\documentclass[12pt, A4paper]{article}
\usepackage[utf8]{inputenc}
\usepackage[colorinlistoftodos,prependcaption,textsize=small]{todonotes}
\usepackage{authblk}
\usepackage{graphicx}
\usepackage{amsmath}
\usepackage[finalnew]{trackchanges}

\newcommand{\beginsupplement}{%
        \setcounter{table}{0}
        \renewcommand{\thetable}{S\arabic{table}}%
        \setcounter{section}{0}
        \renewcommand{\thesection}{S\arabic{section}}%
        \setcounter{figure}{0}
        \renewcommand{\thefigure}{S\arabic{figure}}%
     }

\title{Tricritical fluctuations and elastic properties of the Ising antiferromagnet UIrSi$_3$}
\author[*]{T.N.~Haidamak, J.~Valenta, J.~Prchal, M.~Vališka, J.~Pospíšil, V.~Sechovský, J.~Prokleška}
\affil[*]{Charles University, Faculty of Mathematics and Physicss, Department of Condensed Matter Physics, Ke~Karlovu~5, Prague, Czech Republic}
\author[**,***]{A.A.~Zvyagin}
\affil[**]{B.~Verkin Institute for Low Temperature Physics and Engineering of the National Academy of Sciences of Ukraine,  47, Nauky ave., Kharkiv, 61103, Ukraine}
\affil[***]{V.N.Karazin Kharkiv National University, 4, Svobody sq., Kharkiv, 61022, Ukraine}
\author[****]{F.~Honda}
\affil[****]{Central Institute of Radioisotope Science and Safety Management Kyushu University Motooka 744, Fukuoka-Nishi, Fukuoka 819-0395, Japan}
\date{\change{Sept 2021}{Apr 2022}}

\begin{document}

\maketitle

\subsection*{Abstract}

Elastic constants, thermal expansion, magnetostriction and heat capacity measurements were performed with and without applied magnetic field on a single crystal of UIrSi$_3$. The elastic properties were interpreted within the theory of the strain-exchange effect. The exchange-striction model together with the Ising model for the behavior of magnetic localized 5$f$ electrons and itinerant electrons of uranium correctly reproduces the main features of our magneto-acoustic experiments in UIrSi$_3$. Data on thermal expansion and magnetostriction confirm the conclusion that the dominant contribution of the measured temperature and field change in the speed of sound comes from the change in the elastic modulus itself. Based on the analysis of heat capacity measurements in the magnetic field, we explain the significant anomalies at the second-order branch of the phase transition boundary as a manifestation of the tricritical fluctuations, dominating the region below the tricritical point. The outcome confirms the 3D Ising model, at zero and low magnetic fields, with a crossover to the mean-field tricritical behavior at fields close to the tricritical point, where tricritical fluctuations dominate the temperature evolution of the given property.

\section{Introduction}

The term "elastic properties” usually refers to the properties following the Hook’s law, generally written as $\sigma_{ij} = c_{ijkl} \epsilon_{kl}$, where $\sigma_{ij}$ and $\epsilon_{kl}$ are the stress and strain tensor, respectively, and $c_{ijkl}$ is the elastic constant tensor. Together with the isothermal compressibility, the $\epsilon_{kl}$ are related to the second derivatives of the Gibbs free energy, similarly to the heat capacity, and carry important information about the system. The magnetostructural coupling reflecting the interplay between the spin and lattice degrees of freedom makes the elastic properties useful for investigating thermodynamic phenomena in magnetic materials. 

The physics of uranium compounds has been studied intensively for several decades, revealing numerous exotic physical properties such as a variety of magnetic structures, heavy electrons and their superconductivity, coexistence of magnetism and superconductivity, hidden order, etc. The 5$f$- electrons of U ions play a key role in the emergence of these exciting phenomena. Contrary to the 4$f$-electron orbitals, which are deeply buried in the core electron density of lanthanide ions, the uranium spatially more extended 5$f$-electron wave functions interact with the overlapping 5$f$-orbitals of neighboring U ions (5$f$-5$f$  overlap) as well as with the valence electron orbitals of ligands (5$f$-ligand hybridization) \cite{chapter,Koelling85}. Consequently, the 5$f$‑electron wave functions in U intermetallics lose, to a considerable extent, their atomic character and the U magnetic moment is often found substantially reduced in comparison with the free-ion U moments of U$^{4+}$ or U$^{3+}$. The large 5$f$-5$f$  overlap by rule prevents the formation of a rigid atomic 5$f$-electron magnetic moment in materials in which the distance of nearest-neighbor U atoms is smaller than the Hill limit (340-360 pm)~\cite{Hill}. The 5$f$-ligand hybridization has more subtle effects on magnetism which show up in the lower U-content compounds where the ligands surrounding U ions prevent the direct U-U bonds \cite{chapter, Koelling85}. The direct overlap of 5$f$  U-wave functions is responsible for the direct 5$f$-5$f$  exchange interaction, while 5$f$-ligand hybridization mediates the indirect exchange interaction between U ion moments adjacent to the involved ligand. The strong spin-orbit interaction in heavy uranium ions causes even strongly delocalized 5$f$-electrons to carry a significant orbital magnetic moment, which dominates the spin moment.

The orbital polarization gives rise to a huge magnetocrystalline anisotropy which seems to be inherent to uranium magnetics. Contrary to the single-ion anisotropy observed in 4$f$-electron lanthanide compounds, the strong interaction of the spatially extended U 5$f$-orbitals with surrounding ligands in the crystal and participation of 5$f$  electrons in bonding~\cite{Smith83, Eriksson91} implies an essentially different mechanism of magnetocrystalline anisotropy. The anisotropy of the bonding and 5$f$–ligand hybridization assisted by the strong spin-orbit interaction are the key ingredients of the inter-ion anisotropy of U magnetics. 

The systematic occurrence of particular anisotropy types related to the layout of the U ions in a crystal lattice suggests that the easy magnetization direction is perpendicular to the planes or chains comprising the direct U-U bonds~\cite{chapter,Sechovsky94}. 

Studies of intermetallic compounds crystallizing in a non-centro\-symmetric crystal structure have been impressive in the last two decades. The popularity of these materials, especially the R$TX_3$ compounds, is associated with the observation of unconventional superconductivity, high critical fields, vibron states, quantum criticality, etc. \cite{Bauer2009, Pfleiderer2009,Bauer2012, Kimura2012, Kimura2005, Sugitani2006, Klicpera2017}. The connecting attribute of these intermetallic compounds is the 4$f$-electrons in lanthanide compounds. Only two U compounds crystallizing in the tetragonal non-centrosymmetric crystal structure, UIrSi$_3$ and UNiGa$_3$, are reported in the literature. Both compounds have been studied as polycrystalline samples exhibiting antiferromagnetic (AFM) order below 42~K (UIrSi$_3$) \cite{Buffat1986} and 39~K (UNiGa$_3$) \cite{Takabatake1993}, respectively. 

Recently, we have studied in detail the magnetic, thermal, and transport properties of UIrSi$_3$ together with $ab$-initio calculation of the ground state \cite{Val,Val2} revealing the rich physics of the compound. However, the key question about the relation between the magnetic and crystal structure remained unanswered. In this contribution, we present the elastic properties determined by the measurements of elastic constants, thermal expansion and magnetostriction followed by interpretation within the theory of the strain-exchange effect. Based on the analysis of heat capacity measurements in the magnetic field, we explain the significant anomalies at the second-order branch\add{, i.e. above $T_{\rm tc}$ and below $H_{\rm tc}$, cf Fig.}~\ref{MPD} of the phase transition boundary as a manifestation of the tricritical fluctuation, dominating the region below the tricritical point.

The outcome is discussed in the context of the thermodynamics of the general antiferromagnetic system, more specifically, the behavior along the phase border between an antiferromagnetic ground state and a field polarized ferromagnetic phase.

\section{Experimental}

For the measurements, we used pieces of a single crystal grown by the floating zone melting method in a commercial four-mirror optical furnace with halogen lamps, each  1kW (model FZ-T-4000-VPM-PC, Crystal Systems Corp., Japan), 
\change{originated from the same crystal as in}{being part of} 
 the previous studies  \cite{Val,Val2}. A polycrystalline precursor was synthesized by arc-melting from stoichiometric amounts of the pure elements U  (3N, further treated by Solid-State Electrotransport \cite{sse1,sse2}), Ir (4N), and Si (6N) in Ar (6N) protective atmosphere. The growth process was performed in an Ar (6N) flow of 0.25~l/min and pressure of 2~bar. The pulling speed was very low, only 0.5~mm/h. A large single crystal of the cylindrical shape with length 50~mm and diameter of 4~mm was obtained. The details of the preparation are discussed in \cite{Val2}. The high quality and orientation of the single crystal were verified by the Laue method (\add{see~}Fig. 1\add{~in~{\cite{Val2}}}). The stoichiometric composition was confirmed by scanning electron microscopy (SEM) using a TescanMira I LMH system equipped with an energy-dispersive X-ray detector (EDX) Bruker AXS. The analysis revealed a single-phase single crystal of 1\add{.0(1):}1\add{.0(1)}:3\add{.0(1)} composition. 

A single sample was cut \add{(by wire saw, South Bay Technology 810CE)} out of an oriented crystal with a shape allowing all necessary geometries for both thermal expansion and ultrasonic measurements. The resulting parallelepiped has planes perpendicular to all key directions. Sample thicknesses in the given directions are  1.57~mm for [100]\add{ (a-axis)}, 1.54~mm for [110]\add{ (basal plane diagonal)} and  2.45~mm for [001]\add{ (c-axis)}. Flat parallel polished planes allowed us to measure temperature and magnetic field dependences of the elastic \change{ultrasound wave}{moduli} with \add{given }{\bf q}-wave vector and {\bf u}-polarization.
Due to the known anisotropy of the compound~\cite{Val, Val2} the magnetic field was applied only along the $c$-direction.

For the measurement of compressibility, we used a set of strain gauges (Kyowa Electronic Instruments, type KFL-05-120-C1-11) installed to the clamp pressure cell. High-resolution length changes along the $a$- and $c$-direction were measured using a miniature capacitance dilatometer \cite{Rotter2007} connected to the AH2500A capacitance bridge. The $C_{11}$ and $C_{33}$ elastic moduli were measured using the Ultrasonic option (Quantum Design)\add{ employing phase comparison method for the determination of the acoustic velocity change. The measurements were done in transmission geometry, with LiNbO$_3$ transducers glued to the sample with thiokol LP032}. For the measurement of the elastic moduli, several frequencies were tested in the 17-110~MHz range, no frequency dependence was observable. Both instruments were used in a Physical Property Measurement System (Quantum Design) cryostat for temperature and magnetic field control. The rate of temperature change was typically 20-50 mK/min to keep the sample properly thermalized.

\section{Results}

\begin{figure}
\begin{center}
\includegraphics[width=0.49\textwidth]{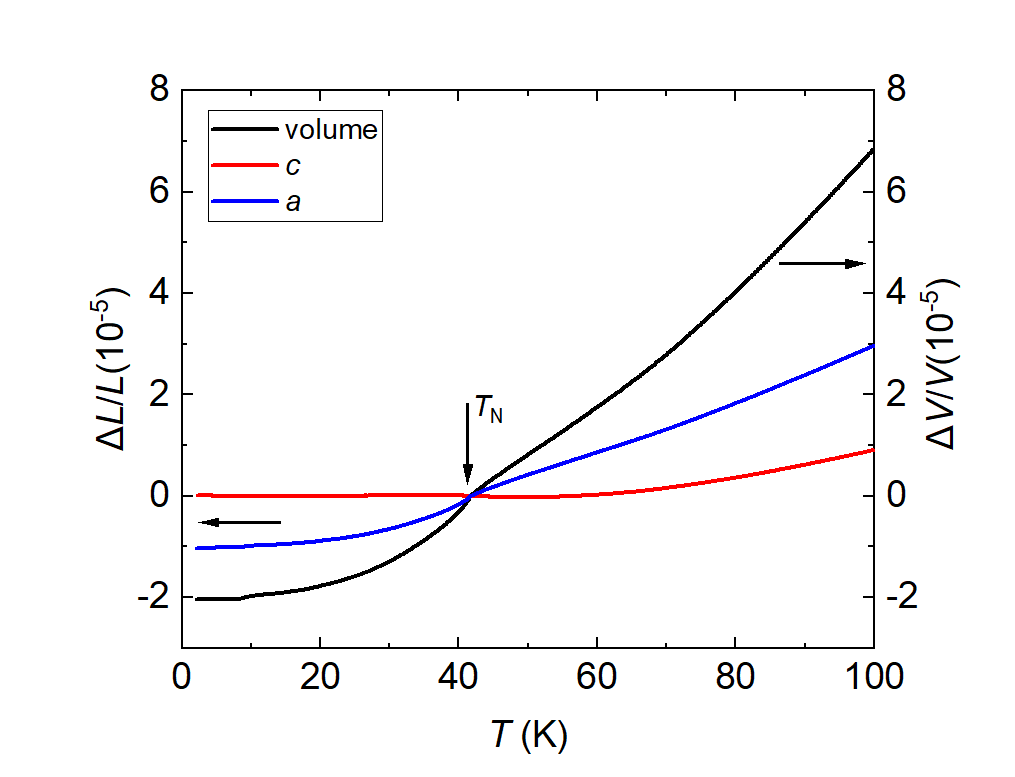}
\includegraphics[width=0.47\textwidth]{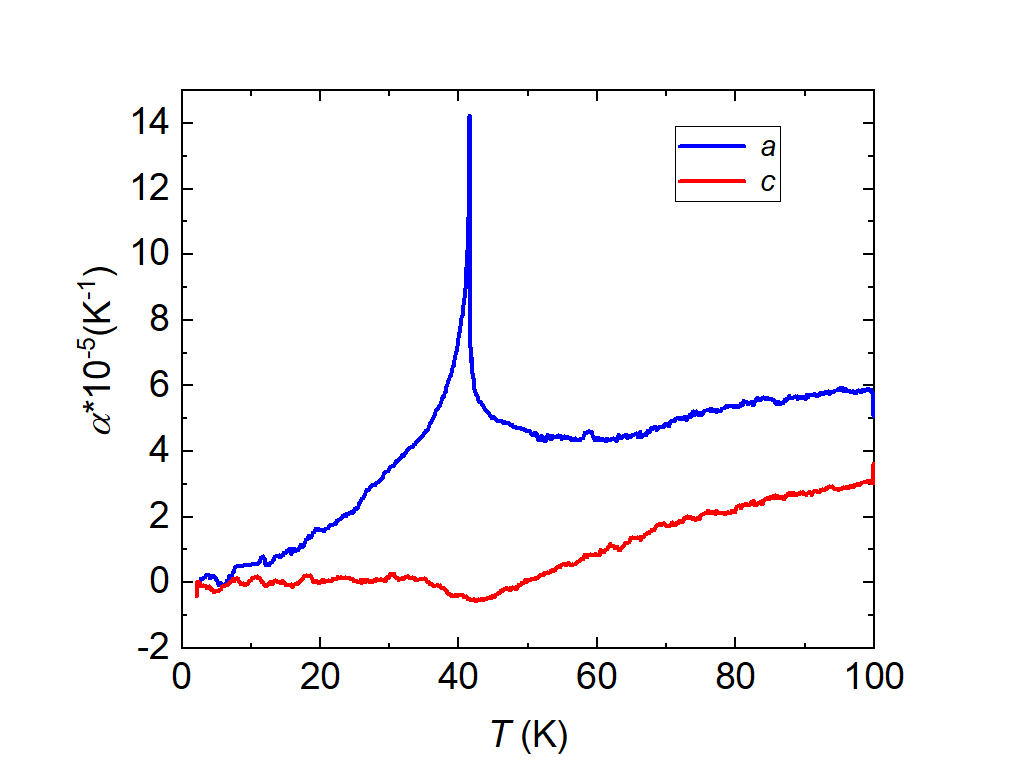}
\end{center}
\caption{Thermal expansion and thermal expansivity as a function of temperature measured along main crystallographic directions.}
\label{figTE}
\end{figure}

The thermal expansion measured along the $a$- and the $c$-axis is displayed in Fig.~\ref{figTE} and shows large anisotropy in agreement with the previous results~\cite{Val,Val2}. Upon cooling, the thermal expansion along the $a$-axis continuously decreases down to the AFM ordering temperature $T_{\rm N}$ where it exhibits an evident change of curvature. On the contrary, the thermal expansion along the $c$-axis decreases at a much lower rate, exhibiting an inflection point at $T_{\rm N}$. Further cooling does not lead to an appreciable change of the $c$-axis thermal expansion. In fact, the lattice parameter $c$ remains practically constant at temperatures below about 25-30 K. The volume change derived from the thermal expansion results along the $a$-axis and $c$-axis ($\Delta V/V=2\Delta a/a + \Delta c/c$), exhibits a pronounced change of curvature at $T_{\rm N}$ being dominated by the $a$-axis contribution.

Fig.~\ref{figTE} shows the temperature dependence of the thermal expansion coefficients. As expected, the magnetic ordering manifested as a peak-shape anomaly is much more pronounced in the $a$-axis thermal-expansion coefficient ($\alpha_a$) than in the $c$-axis thermal-expansion coefficient ($\alpha_c$). Knowing the step of the thermal-expansion coefficients at $T_{\rm N}$ together with the step of the specific heat divided by temperature ($\Delta C_p /T$) at the magnetic ordering temperature~\cite{Val}, the pressure dependences can be calculated via the thermodynamic Ehrenfest relation\change{s.}{ for second order phase transitions}
\begin{equation}
    \frac{{\rm d}T_{\rm N}}{{\rm d}p} = V_{\rm m} \frac{\Delta\alpha}{\Delta C_p /T}.
\end{equation}
The anisotropic pressure dependence can be calculated from respective $\Delta\alpha_a$ and $\Delta\alpha_c$ leading to a quite high hydrostatic-pressure dependence of $T_{\rm N}$ amounting to ${\rm d}T_{\rm N} /{\rm d}p = 1.93(5)$K/GPa.

\begin{figure}
\begin{center}
\includegraphics[width=0.49\textwidth]{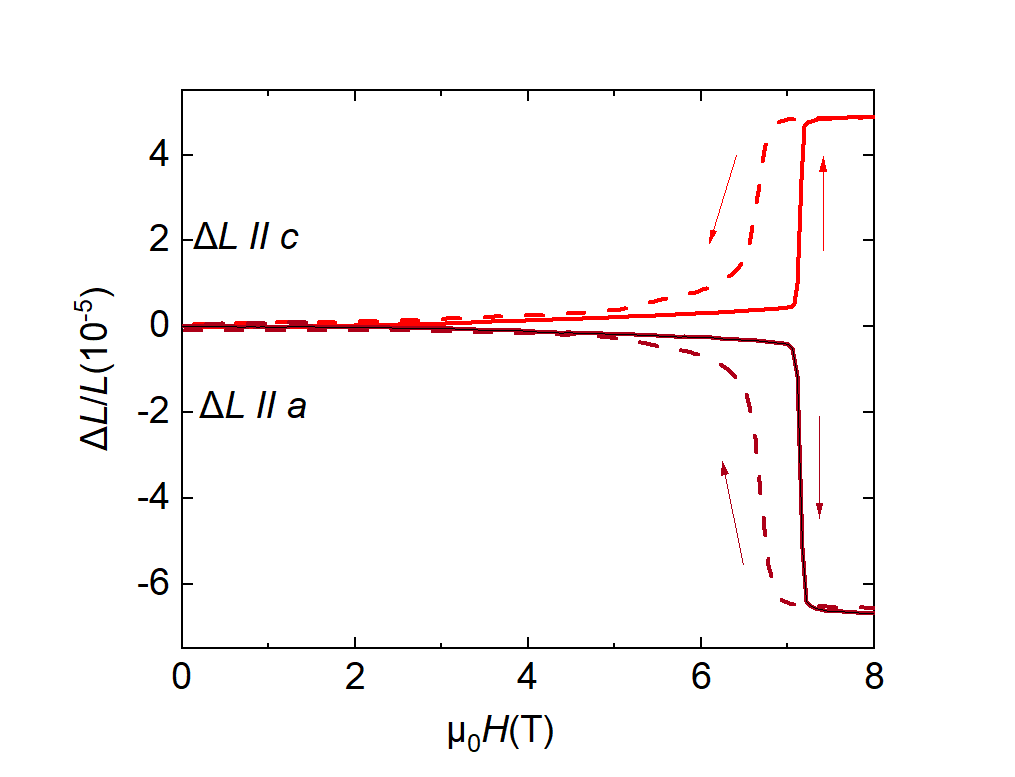}
\includegraphics[width=0.485\textwidth]{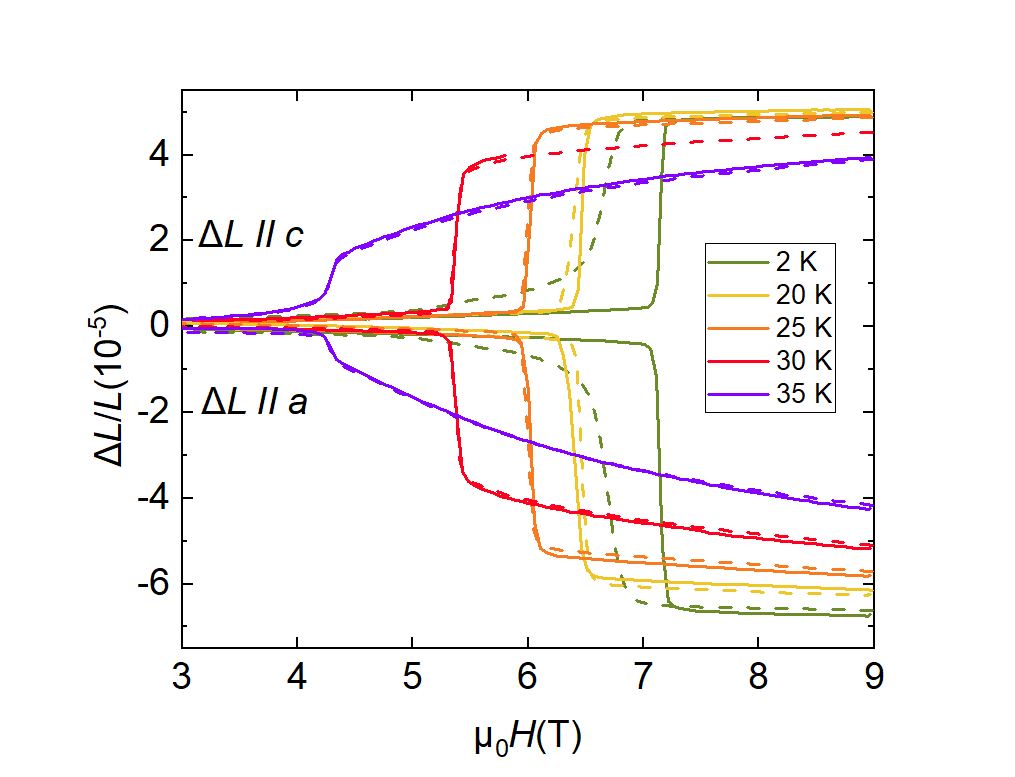}
\end{center}
\caption{Longitudinal and transversal magnetostriction measured for field applied along the $c$-axis at 2K (left panel, the arrows show the sweep direction of the magnetic field) and at several fixed temperatures (right panel).}
\label{figMS}
\end{figure}

The magnetostriction isotherms were measured along the $a$- and the $c$-axis in a magnetic field applied along the $c$-axis. As can be seen in Fig.~\ref{figMS} at low temperatures, step-like changes of the sample dimensions are observed at a characteristic field. For a field sweep up, the magnetostriction shows a sharp expansion along the $c$-axis and a simultaneous sharp contraction along the $a$-axis (yielding volume reduction). The anomalies exhibit hysteretic behavior characteristic of a first-order phase transition. The asymmetric hysteresis observed in the measurements of other properties \cite{Val,Val2} is also observed in the magnetostriction measurements. The values of the characteristic field of magnetostriction anomalies coincide with metamagnetic transition fields in the magnetic phase diagram of UIrSi$_3$ presented earlier \cite{Val,Val2}. When increasing temperature, the change of the order of the transition at the tricritical point ($T_{\rm tc}=28$~K) is observed. The step-like anomaly smears out at higher temperatures up to the $T_{\rm N}$ indicating a continuous (second-order) phase transition. As with the thermal expansion along the $c$-axis, the magnetostriction along the $c$-axis in the high field is almost temperature-independent at temperatures up to $T_{\rm tc}$ in contrast to the considerable temperature dependence of both properties along the $a$-axis. 

The magnetostriction isotherms qualitatively resemble the magnetization isotherms. At temperatures above $T_{\rm tc}$, the height of the expansion step gradually decreases with increasing temperature, and the field hysteresis of the \change{MT}{metamagnetic transition} vanishes with increasing temperature for both magnetostriction measurements in the same way.

Knowing the temperature and field dependences of relative changes of lattice parameters in main directions, it is worth discussing the behavior of tetragonality ($c$/$a$ ratio) in the form of its relative change $\frac{\Delta c/a}{c/a}=(\Delta c/c - \Delta a/a$). With decreasing temperature, the tetragonality monotonously decreases, with the small dip in the vicinity of the Néel temperature. Contrary to this, there is a large step-like change at the metamagnetic transition amounting to $1.1\cdot 10^{-4}$\add{ (see Fig.~}\ref{tetragonality}). This corroborates the scenario with the compensated AF ground state at low temperatures and fields with the spin-flip metamagnetic transition at $B_c$.

\begin{figure}
\begin{center}
\includegraphics[width=0.475\textwidth]{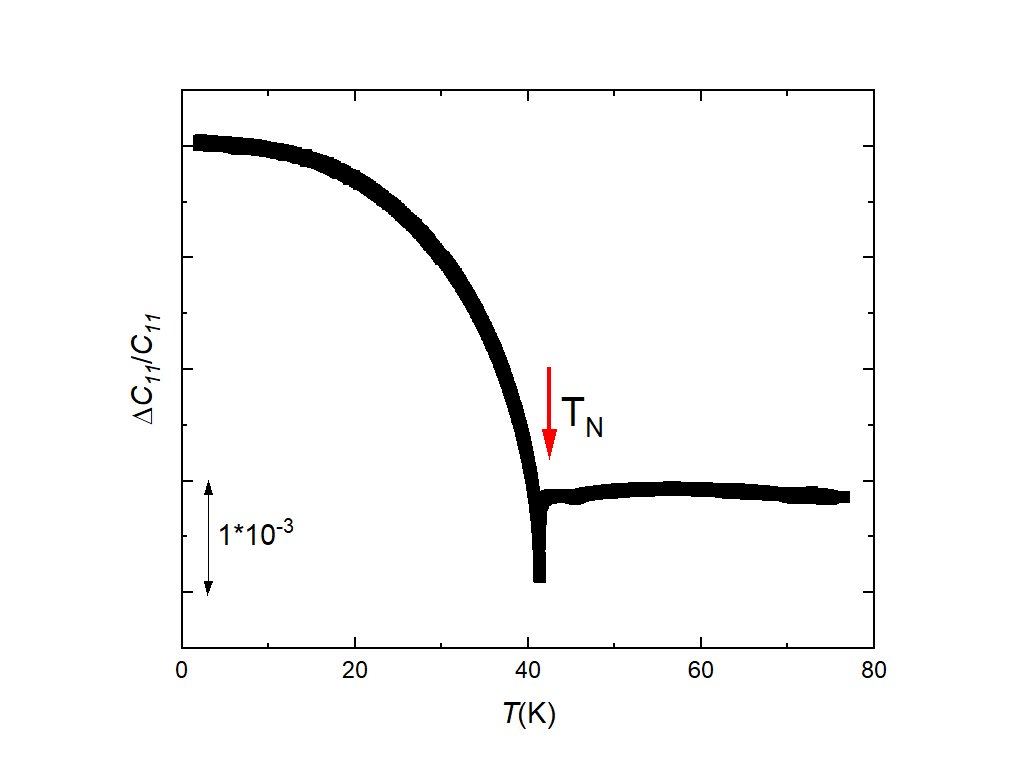}
\includegraphics[width=0.49\textwidth]{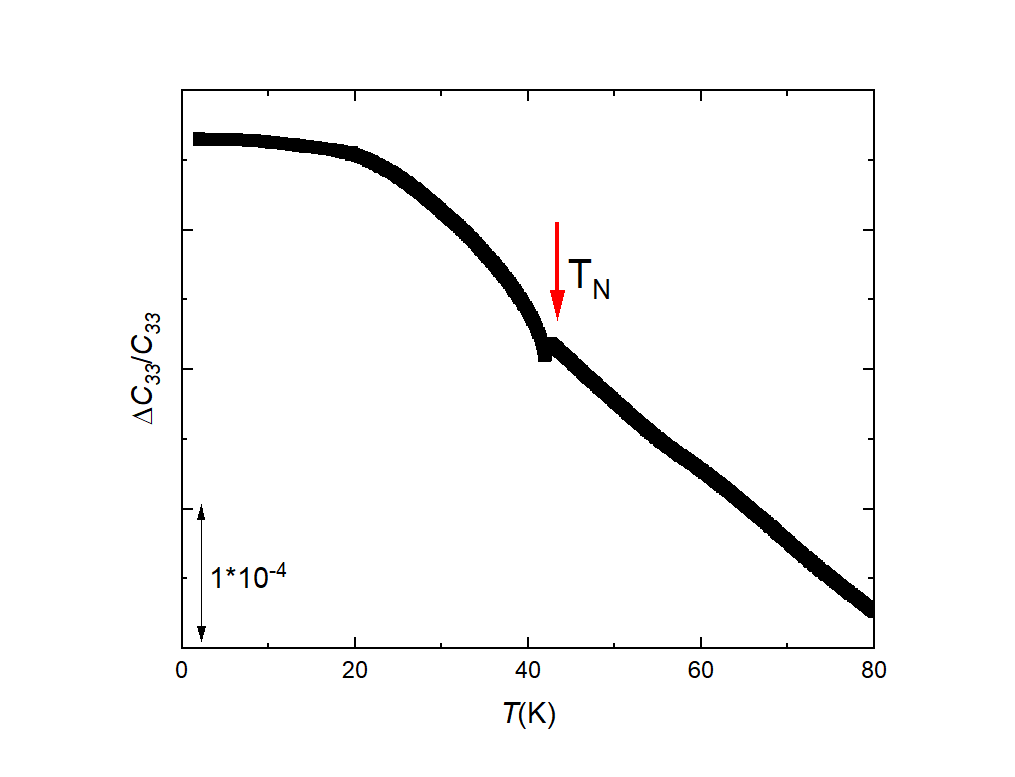}
\end{center}
\caption{Temperature dependencies of $C_{11}$ and $C_{33}$ moduli for zero magnetic field.}
\label{figUZ_T}
\end{figure}

Whereas thermal expansion data bear information on the change of the absolute dimensions of the sample, the measurement of the elastic constants brings complementary information on the elasticity of the material under the investigation.
The zero-field-cooled (ZFC) temperature dependences of frequency of ultrasound signal which propagates in an elastic medium were measured along the $c$–and $a$–axis. The $C_{11}$ and $C_{33}$ moduli temperature dependences with elastic wave propagation along the $a$-axis (${\bf q}\parallel $ \change{$x$}{$[100]$}, ${\bf u} \parallel $ \change{$x$}{$[100]$}) and $c$-axis (${\bf q}\parallel  $ \change{$z$}{$[001]$}, ${\bf u} \parallel $ \change{$z$}{$[001]$}), respectively, were obtained (Fig.~\ref{figUZ_T}). For both moduli, the Néel temperature is visible as a dip in the temperature dependence.

\begin{figure}
\begin{center}
\includegraphics[width=0.485\textwidth]{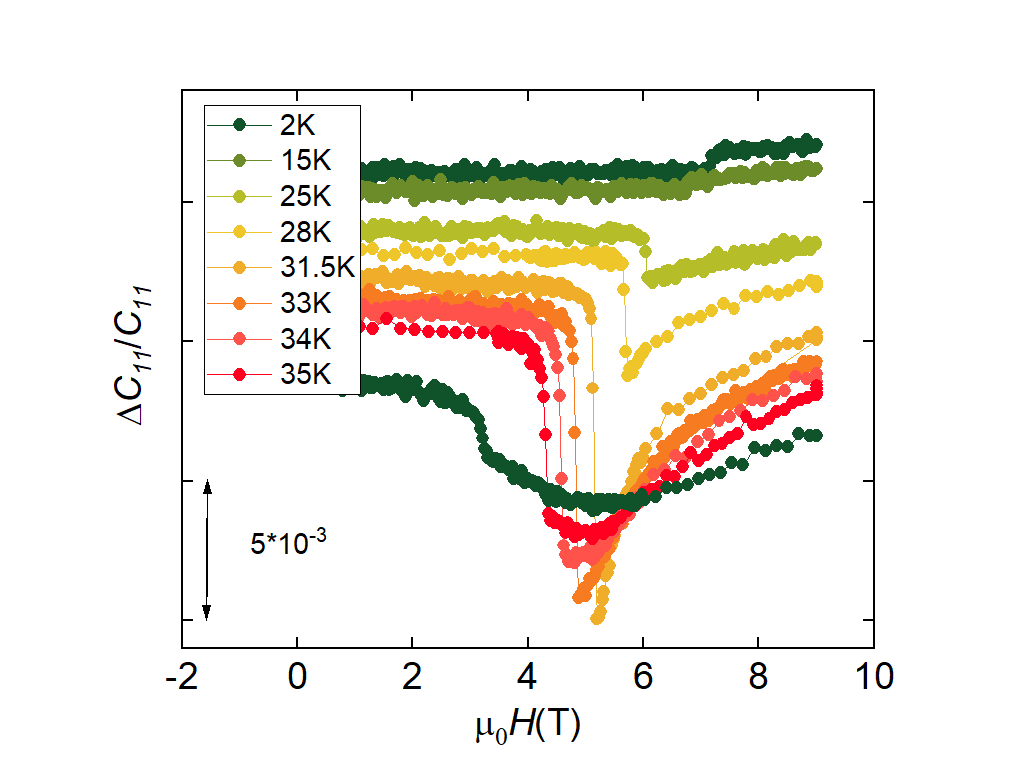}
\includegraphics[width=0.49\textwidth]{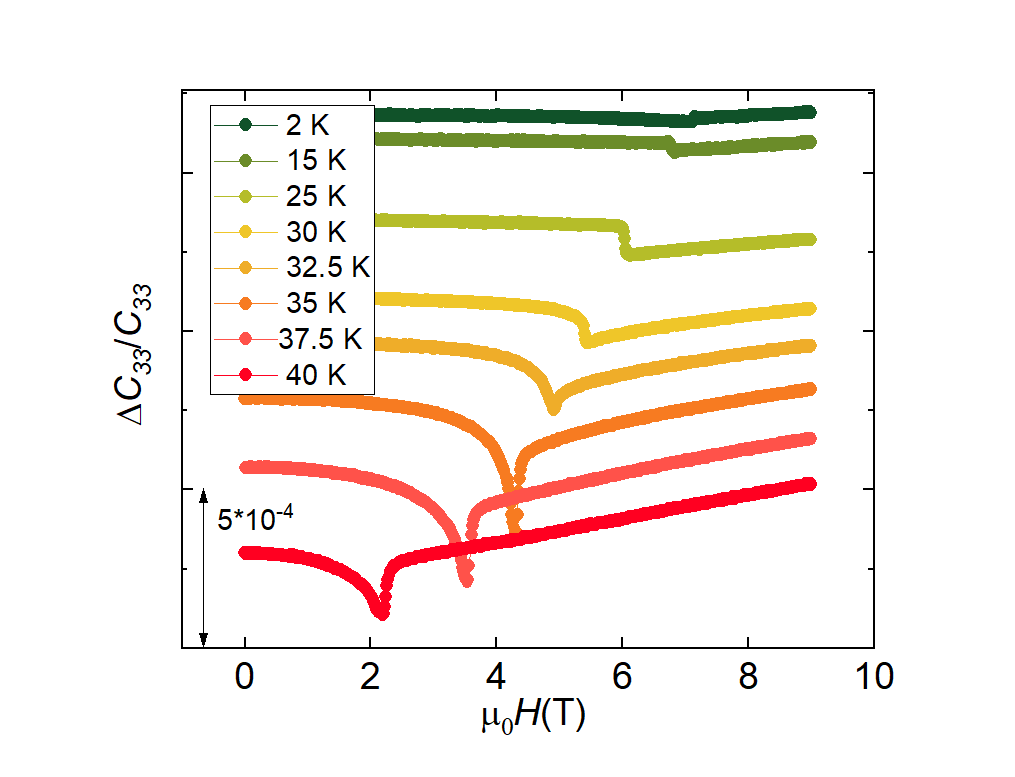}

\end{center}
\caption{Isothermal field dependencies of $C_{11}$ and $C_{33}$ moduli for field applied along the $c$-axis, the curves are vertically shifted for better readability. \add{Selected data are presented, see Fig~}\ref{figUZ_B_all}\add{ for complete data sets.}
}
\label{figUZ_B}
\end{figure}

The field dependencies of \add{the longitudinal} elastic moduli are presented in Figs.~\ref{figUZ_B} \add{and}~\ref{figUZ_B_all}. Contrary to the magnetostriction, we see minimum change across the metamagnetic transition on the observed properties at low temperatures\remove{ and the change of the length of the sample in the given direction is the presumable cause of the anomalies in the measured elastic moduli}. On the other hand --- at elevated temperatures, the change of the measured signal at the Néel temperature is dominated by the change of the elastic moduli\remove{ indicating the strong involvement of phonons}.

Following the transition line across the $B-T$ space, we see a non-monotonous effect across the transition --- at low temperatures and high magnetic fields the effect is rather small, whereas with decreasing field (increasing temperature) the transition signature is enhanced with a maximum \change{around 35 K}{between 28 K and 36K, see Fig.}~\ref{transline}. With further increasing temperature towards the $T_{\rm N}$ the transition signature is suppressed. The downturn below the metamagnetic transition at high temperatures reflects the increased magnetization due to the onset of the spin-flip process.

\add{For completeness, we present (see Fig.}~\ref{figUZ_TM}\add{) the transverse elastic moduli $C_{44}$ (${\bf q}\parallel [100]$, ${\bf u} \parallel [001]$) and $(C_{11}-C_{12})/2$ (${\bf q}\parallel [110]$, ${\bf u} \parallel [1\bar 1 0]$) in Supplemental Materials. The effects observed are similar to longitudinal ones with one notable exception --- the $(C_{11}-C_{12})/2$ show monotonous behaviour across the whole transition line. }

\section{Discussion}

\subsection*{Magneto-acoustic characteristics in the antiferromagnetic Ising  model}

The \add{magnetoelastic interaction influences the} renormalization of sound characteristics in magnetic systems \remove{can be caused} by two \change{reasons}{factors}. \change{S}{Firstly, s}ound wave \change{can }{parameters, such as the sound velocity and attenuation, are sensitive to} change \add{of} positions of nonmagnetic ions, ligands, surrounding magnetic ions. As a result, the crystalline electric field of ligands is affected. Due to the strong spin-orbit interaction, the crystalline electric field renormalizes the single-ion magnetic anisotropy of magnetic ions and the effective $g$-factors of magnetic ions. This is why sound waves can  \change{change}{be used for probing of} magnetically anisotropic properties of single magnetic ions. This effect has a relativistic nature. Such a strain-single ion effect exists at any temperature lower than, e.g., the characteristic energy of the single-ion magnetic anisotropy. 

On the other hand, sound waves \change{can change}{are sensitive to} the positions of magnetic ions and/or positions of nonmagnetic ions, involved in the indirect exchange coupling. As a result, sound waves \change{change}{probe} the effective exchange interactions between magnetic ions. This effect is more pronounced than the strain-single-ion coupling discussed above since the inter-ionic exchange interactions determine predominantly magnetic phase transitions in ordered magnets. 

Consider such an exchange-striction mechanism to describe our experimental observations in UIrSi$_3$. According to \cite{TM}, the exchange-striction coupling in magnetic systems produces the renormalization of the velocity of the sound wave, proportional to spin-spin correlation functions. Those correlation functions can be approximated by a combination of the magnetization and the magnetic susceptibility of the system \cite{TM}. References~\cite{Erf2014,Zher2015,And2017,UN,Nom2020} present good agreements between magneto-acoustic experiments and the theory of the strain-exchange effect for many magnetic systems even if only the homogeneous part of the magnetic susceptibility is taken into account. According to that simplified theory, the renormalization of the sound velocity, $\Delta v$, due to the exchange-striction coupling for the homogeneous contribution from the magnetic susceptibility can be presented as 
\begin{eqnarray}
&&\frac {\Delta v}{v} \approx  - \frac {v}{\rho V \omega^2\mu^4} 
\biggl[|g(0)|^2(2M^2\chi + \nonumber \\
&&k_BT\chi^2) + h(0)\mu^2(M^2 +k_BT\chi)\biggr], \
\end{eqnarray}
where $V$ is the volume of the crystal, $\mu =g\mu_B$ is the effective magneton per magnetic ion ($g$ is the effective $g$-factor, and $\mu_B$ is Bohr's magneton), $\chi$ is the magnetic susceptibility, $v$ is the velocity of sound, $\omega$ is the frequency, $k_B$ is the Botzmann constant, and $T$ is the temperature. The magnetoelastic coefficients can be written as \cite{TM}
\begin{eqnarray}
&&h({\bf q}) =\sum_j e^{-i{\bf q}{\bf R}_{ji}}\left[ 1-\cos ({\bf k}{\bf R}_{ji})\right] \times \nonumber \\ 
&&\times ({\bf u}_{\bf k}\cdot {\bf u}_{-{\bf k}})\frac {\partial^2 J_{ij}^{\beta,\beta '}}{\partial {\bf R}_i \partial {\bf R}_j}, \ \nonumber \\ 
&&g({\bf q}) = \sum_j e^{i{\bf q}{\bf R}_{ji}}\left( e^{i{\bf k}{\bf R}_{ji}} -1\right) {\bf u}_{\bf k} \frac {\partial J_{ij}^{\beta,\beta '}}{\partial {\bf R}_i} 
\end{eqnarray}
(taken at ${\bf q}=0$), where ${\bf R}_{ji} = {\bf R}_j -{\bf R}_i$, ${\bf R}_j$ is the position vector of the $j$-th site of the magnetic ion, and $J_{ij}^{\beta,\beta '}$ ($\beta,\beta ' = x,y,z$) are the exchange couplings between magnetic ions on the $i$-th and $j$-th site, ${\bf k}$ and ${\bf u}_{\bf k}$ are the wave vector and the polarization of the sound wave, respectively. Usually those magnetoelastic coefficients are used as fitting parameters.   

Now, our goal is to calculate the magnetic field and temperature behavior of the magnetization and the magnetic susceptibility of UIrSi$_3$. When doing that, we have to take into account the dual nature of the 5$f$ electrons. Localized states of U ions lying near the Fermi energy can hybridize with conduction electrons and, thus, experience a weak dispersion, enhancing the density of states at the Fermi surface. On the other hand, only part of the 5$f$ electrons of U ions may become itinerant, while the rest remains localized in the vicinity of the Fermi surface. It was considered in the so-called dual-nature model~\cite{Zv2001, Zw2001, Zw2003, Zv2003, Pol2006} with competing localized and delocalized 5$f$ electrons for both exact and perturbative theoretical approaches. Such a dual model explains the coexistence of the magnetic and conducting properties of several U compounds. 

First, let us estimate the contribution to the magnetization and magnetic susceptibility from the conduction (band) part of 5$f$ electrons. The magnetic (Pauli) susceptibility of the conduction electrons (assumed to be noninteracting in our estimates) can be obtained from the value of the Sommerfeld coefficient, 31~mJ mol$^{-1}$ K$^{-2}$, known from the linear in $T$ contribution to the specific heat~\cite{Val}. It gives $\chi_{it} \approx 10^{-8}$m$^3$mol$^{-1}$. The contribution from conduction electrons to the magnetic characteristics is essential in UIrSi$_3$ at low values of the field at low temperatures, confirming the dual nature of the 5$f$ electrons in that compound. On the other hand, itinerant electrons themselves cannot explain the observed spin-flop-like phase transition, which is related to the localized magnetic moments (see below).

To describe the magnetic properties of the localized electrons of U ions in UIrSi$_3$ we can consider the antiferromagnetic Ising model with the easy-axis direction along $c$-axis. It is known that the antiferromagnetic Ising model describes well the metamagnetic phase transition \cite{Mot}, which takes place in UIrSi$_3$ \cite{Val}. In our analysis, we follow Refs.~\cite{Mot,BVY}. Let us start with the Ising Hamiltonian
\begin{equation}
{\cal H} = \sum_{l,m}J_{(l,m)}S_lS_m -\mu H \sum_l S_l \ , 
\label{H}
\end{equation}
where the summation is over the sites of the magnetic lattice, $J_{(l,m)}$ are the exchange integrals, $H$ is the external magnetic field, and $S_l$ are the  operators of the Ising components of the site spins (which can be equal to $\pm 1$). Suppose that the main interaction is antiferromagnetic, and, therefore, we can divide all magnetic moments of localized electrons of U ions into two sublattices. In the mean field approximation we can write the free energy of the considered Hamiltonian per site as
\begin{align}
F_M &= -\mu H(M_1+M_2) -\frac{J_1}{2}[M_1^2+M_2^2] + \nonumber \\
&J_2M_1M_2 - T \left( {\cal S}[M_1] + {\cal S}[M_2]\right)  \ , 
\label{free1}
\end{align}
where $M_{1,2} \equiv \langle S_{1,2}\rangle$ are the average values of operators of magnetic moments of each of magnetic sublattices, and $J_{1,2} \ge 0$ are the effective exchange parameters, which describe the interaction between spins, belonging to the same sublattice, and belonging to different sublattices, respectively. Here we denote ${\cal S}(x) =k_B[\ln 2 -(1/2)(1+x)\ln(1+x) -(1/2)(1-x)\ln (1-x)]$ the entropy of the Ising spin. Then, let us introduce $M=(M_1+M_2)/2$ and $L=(M_1-M_2)/2$, the magnetization and the order parameter of the antiferromagnetic Ising system (in units of $\mu$). We can find the solution $M_0$ for the magnetization $M$ as a function of $L$ for given values of $J_{1,2}$ for small $L$, the magnetic field and temperature, which satisfies the equation
\begin{equation}
\mu H  = (J_2-J_1)M_0 + k_BT \tanh^{-1} M_0 \ .
\label{M0}
\end{equation}
Then, the magnetic susceptibility is equal to 
\begin{equation}
\chi = \frac{\mu}{(J_2-J_1)(1-M_0^2) +k_BT} \ . 
\end{equation}
Substituting this solution into Eq.~(\ref{free1}), and taking into account that the order parameter $L$ is small in the vicinity of the phase transition to the ordered phase, we get
\begin{align}
F_M &\approx F_0 + \frac {aL^2}{ 2} + \frac {bL^4}{4} +
\frac {cL^6}{ 6} + \frac {dL^8}{ 8} + \dots \ , \label{free2} \\
F_0 &= JM_0^2 - \mu H M_0 - 2k_BT{\cal S}(M_0) \ , \nonumber
\end{align}
with 
\begin{equation}
a= \frac{2k_BT}{(1-M_0^2)} -2J_2 \ , 
\end{equation}
where $M_0$ is the solution of Eq.~(\ref{M0}). Near the point $a=0$ we can approximate $a \approx \alpha \mu (H-H_1)$, where $\alpha = 2(J_1+J_2)\sqrt{1-[k_BT/(J_1+J_2)]}/J_2T$. Other coefficients can be calculated in a similar way \cite{BVY}
\begin{eqnarray}
&&b=\beta k_B(T-T_{tc}) \equiv \frac{2J_1(J_1+J_2)^2k_B(T-T_{tc})}{J_2(k_BT)^2} \ , \nonumber \\
&&c= \frac {4(5J_1-3J_2)(J_1+J_2)^5}{45J_2(k_BT)^4} \ , \nonumber \\
&&d= \frac{64(5J_2-7J_1)(J_1+J_2)^7}{2835 J_2 (k_BT)^6} \ ,  
\end{eqnarray}
where $T_{tc} = (J_1+J_2)(3J_1-J_2)/3k_BJ_1$. The free energy (\ref{free2}) differs from Eq.~(\ref{free1}): It is correct only in the vicinity of the phase transition. In fact, it is the effective energy of the Landau phase transition theory for the microscopic model defined by Eq.~(\ref{H}). 

Then according to the standard Landau theory an ordering takes place at $a=0$, i.e., at $M_0 = \sqrt{1-(T/T_N(0))}$, where $T_N(0) = (J_1+J_2)/k_B$. The paramagnetic phase becomes unstable at the line determined by the equation \add{for illustration, cf.~}\ref{MPD}
\begin{eqnarray}
&&\mu H_1 = (J_2-J_1)\sqrt{1 -\frac{T}{ T_N(0)}} \nonumber \\
&&+k_BT\tanh^{-1}\left[\sqrt{1-\frac{T}{ T_N(0)}}\right] \ . 
\label{para}
\end{eqnarray}
At this line, generally speaking, the second order phase transition between an antiferromagnetically ordered phase and the paramagnetic one takes place (see, however, the low-temperature analysis below). Depending on the values of the coefficients $c$ and $d$, at low temperatures in the Ising model there can exist the first order phase transition with the jump of the magnetization. We know from the experiments on low temperature magnetization in UIrSi$_3$ \cite{Val} that really such a situation takes place: the metamagnetic transition is observed at low temperatures. According to the analysis of the considered model \cite{BVY}, in UIrSi$_3$ the situation is realized with $(3J_2/5) \ge J_1 < J_2$. For that case there exists a tricritical point in the $H$-$T$ phase diagram, at which the higher-temperature second order phase transition line is transformed to the lower-temperature first order phase transition line, cf. \cite{Val}. It happens at the temperature $T_{tc}$ with the critical value of the magnetic field being $H_{tc} = (J_2-J_1)\sqrt{1-T_{tc}/T_N(0)} +k_BT_{tc}\tanh^{-1}[\sqrt{1-(T_{tc}/T_N(0)}]$. According to the theory, for $T< T_{tc}$ in the region of parameters $J_{1,2}$, applicable for UIrSi$_3$, there can exist three critical lines on the $H$-$T$ phase diagram\add{ for illustration, cf.~}\ref{MPD}. At the first line, $H=H_1$, the paramagnetic line becomes unstable. At the second line, $H=H_2$, where
\begin{eqnarray}
&&\mu H_2 = \mu H_1 +\frac{2k_BT_N(0)}{\alpha} \biggl(\frac{c\beta(T-T_{tc})}{6 d T_N(0)} - \nonumber \\
&& \frac{c^3}{27d^2} +\left[ \frac{c^2}{9d^{4/3}} -\frac{\beta(T-T_{tc})}{3d^{1/3}T_N(0)}\right]^{3/2}\biggr)  
\end{eqnarray}
the antiferromagnetic phase becomes unstable for $T < T_{tc}$. The presence of two different lines of stability of two phases is the basic feature of the first order phase transition. In the ground state the first critical line starts from the value $\mu H_1 = J_2-J_1$, and the second line from $\mu H_2 = J_1+J_2$
.
Between those two lines there exists the line of the metamagnetic first-order phase transition, which starts in the ground state from the value $\mu H_{sp} =J_2$. 

If the magnetic field is directed perpendicular to the easy axis $c$, the magnetization is increased smoothly with the growth of the value of the magnetic field. Namely, that situation takes place in UIrSi$_3$ \cite{Val}.  

\begin{figure}
\begin{center}
\includegraphics[scale=.6]{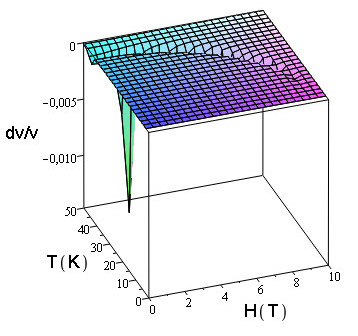}
\end{center}
\caption{(Color online) Calculated relative changes of the sound velocity in UIrSi$_3$ as a function of the temperature and the external magnetic field 
calculated within the strain-exchange model and the Ising model.}
\label{fig1}
\end{figure}

The experiments on the magnetic properties of UIrSi$_3$ imply $T_N(0)(=J_1+J_2)\approx 42$~K. The relative values of $J_1$ and $J_2$ can be estimated from the low-temperature values of the fields of stability of the paramagnetic and antiferromagnetic phases, observed in UIrSi$_3$ in magnetic experiments \cite{Val} and in the present magneto-acoustic ones. Taking into account the above considerations, the results of the approximate numerical calculations in the framework of the Ising model are shown in Fig.~\ref{fig1} for the magnetic field and temperature dependence of the relative change of the sound velocity. Here the linear in $H$ contribution, related to the itinerant $5f$ electrons of U ions, is also included. One can see that the model describes the features of the behavior of magneto-acoustic characteristics of UIrSi$_3$ well. We see that the features in the magnetic field and temperature behavior of $\Delta v/v$ well reproduce the phase $H$-$T$ diagram of UIrSi$_3$ obtained from the magnetic and thermal measurements \cite{Val}, as well as in our magneto-acoustic experiments. However, we cannot speculate about the total quantitative agreement of the results of model calculations and the data of our magneto-acoustic measurements. 
  
In summary, the exchange-striction model together with the Ising model for the behavior of magnetic localized 5$f$ electrons and itinerant electrons of U correctly reproduces the main features of our magneto-acoustic experiments in UIrSi$_3$. 

\subsection*{Evolution of properties along the phase transition line}

The antiferromagnetic ground state is encapsulated by a phase transition line in the $B$--$T$ phase space, separating it from paramagnetic (high temperatures) or field-polarized paramagnetic (low temperature, high fields) states. As reasoned above and observed in earlier publications~\cite{Val,Val2}, the phase transition is of the first-order type at low temperatures, passing the tricritical point and reaching zero field $T_{\rm N}(0)$ via the second-order type branch. Following the transition line, it is notable that the effect on the measured properties in its vicinity is not monotonous, we see that the maximum impact is at temperatures slightly above the $T_{\rm tc}$. In the following, we will discuss possible reasoning behind this phenomenon and its extension to other observables.

\begin{figure}
\begin{center}
\includegraphics[width=0.325\textwidth]{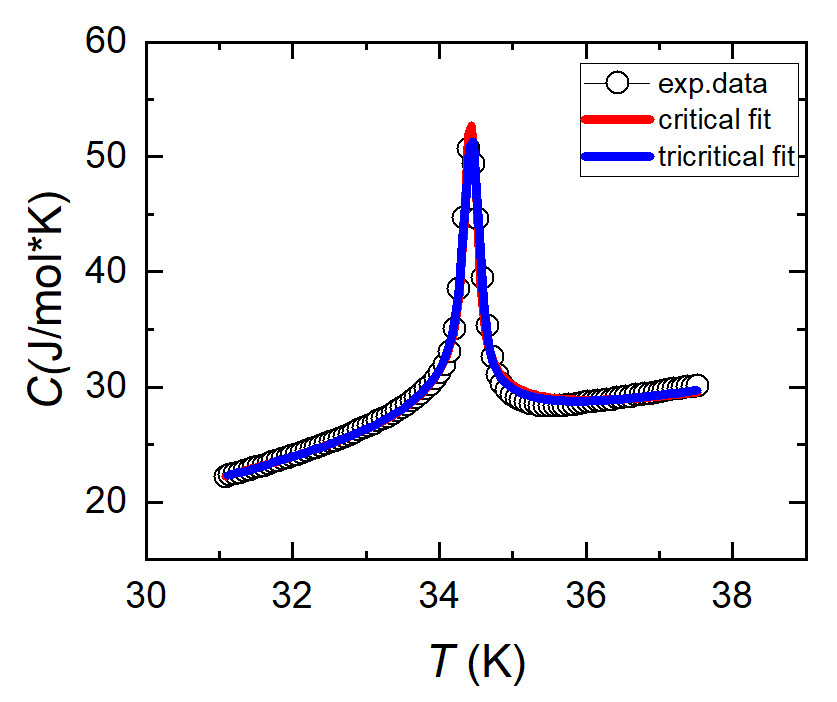}
\includegraphics[width=0.325\textwidth]{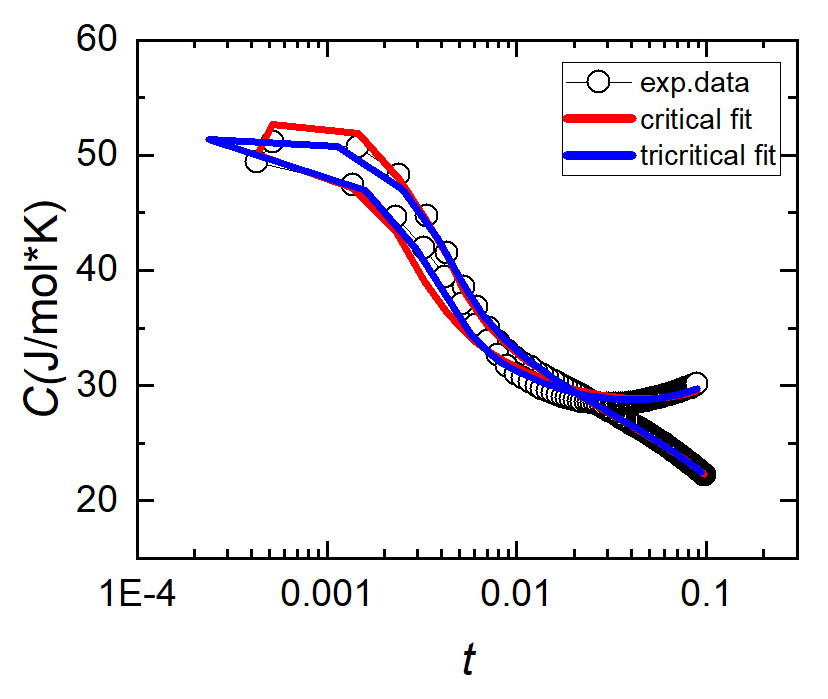}
\includegraphics[width=0.325\textwidth]{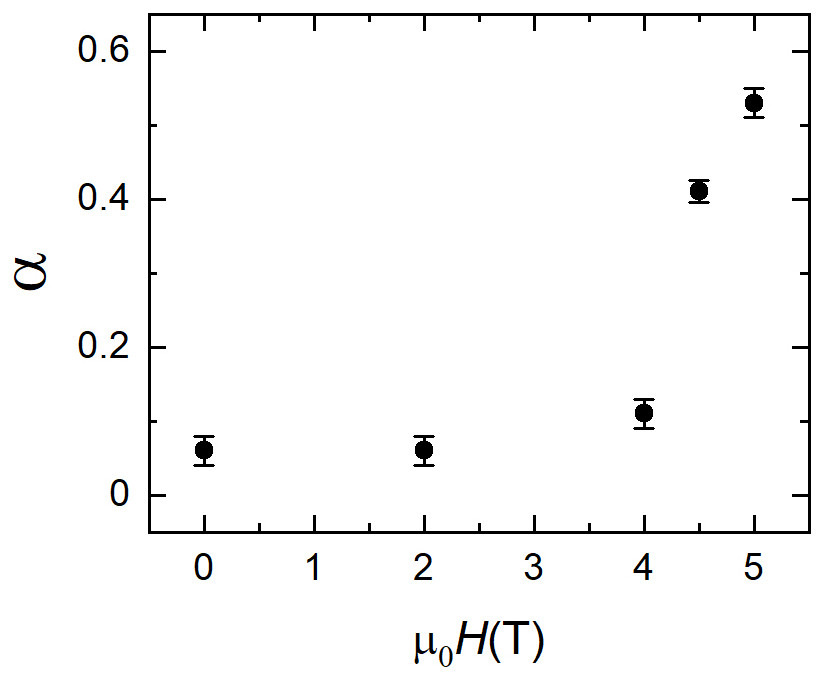}
\end{center}
\caption{Illustration of the fits (tricritical and critical) of the heat capacity data at 4.5T, left in normal scale, middle semilog plot. On the right the field dependence of $\alpha$ coefficient for the critical fit~(\ref{critical}). \add{See Fig.~}\ref{HC}\add{ for original temperature dependencies.}
}
\label{trivscri}
\end{figure}

To aid the discussion, we have repeated earlier measurement~\cite{Val} of heat capacity data for magnetic fields applied along the $c$-axis, this time with the focus on the immediate vicinity of the transition temperature~$T_{\rm N}(B)$. To quantify the evolution in this narrow region we employ the usual description of the critical behaviour compromising the critical contribution \add{($\frac{A^\pm}{\alpha}t^{-\alpha}$, $\alpha$ critical coefficient and $A^\pm$ the respective constants for branches above($+$) and below($-$) ordering temperature)} and the linear background \add{($Et+B$)}
\begin{eqnarray}
\Delta C_{\rm c}(t)=\frac{A^\pm}{\alpha}t^{-\alpha}+Et+B
\label{critical}
\end{eqnarray}
as a function of reduced temperature $t=\frac{\|T-T_{\rm N}\|}{T_{\rm N}}$. The rounding effects were accounted for by a Gaussian distribution of $T_{\rm N}$ (with the outcome of the fit $\delta T_{\rm N}/T_{\rm N}\sim 10^{-3}$). 

The resulting field evolution of the $\alpha$ exponent is shown in Fig.~\ref{trivscri}
accompanied by examples of fit. We see that zero (and low) field exponent is close to the expectation given by the 3D Ising model ($\alpha=0.11$~\cite{CritExp}). This value changes abruptly above 4T to the new value $\alpha\sim 0.5$, indicating the change of the critical behaviour to the mean field tricritical. Indeed, if we perform standard calculations and look for the singular part of the heat capacity $\Delta C= -T\partial^2 F/\partial T^2$ for the free energy~(\ref{free2}) we get (cf., from the point of thermodynamics, similar antiferromagnetic system CsCoBr$_3$~\cite{CsCoBr3}, or related smectic A-C phase transition in liquid crystals~\cite{smectic,smectic2} or spin-Peierls transition in CuGeO$_3$~\cite{CuGeO3})
\begin{eqnarray}
\Delta C_{\rm tc}(t)=C^\pm \left( 1+3\frac{t}{\tau}\right)^{-\frac{1}{2}}
\label{tricritical}
\end{eqnarray}
where $t$ is the reduced temperature defined earlier and $\tau=1-\frac{T_{\rm cr}}{T_{\rm N}}$, with $T_{\rm cr}$ being the  crossover  temperature  from  tricritical  to critical (3D Ising in this case) behavior. Fitting the measured data to the equation~(\ref{tricritical}) lead to improved agreement with the experimental data.

The above-mentioned reasoning indicates the presence of a sizeable sixth-order coefficient in the Landau expansion (\ref{free2}) and explains the notable anomalies beyond the tricritical point, which extends from tricritical point [5.8T, 28K] down to [4T, 36K] along the transition line. This dominance of the asymptotic tricritical behavior is visible not only on the experimental data on the relative sound velocity change (see Fig.~\ref{figUZ_B}\add{ and Fig}~\ref{transline}) and up to a certain extent captured in the calculated relative sound velocity changes~(see Fig.~\ref{fig1}), but it is present also in the temperature and field dependencies of the resistivity and Hall resistivity~\cite{Val2}.

\section*{Conclusions}

We investigated in detail the elastic properties of the noncentrosymmetric tetragonal compound UIrSi$_3$. Our experimental investigation focuses on \change{both static (}{investigation of} thermal expansion\remove{) and dynamic (speed of sound -} elastic coefficients\remove{) properties}. The study is supported by the thermodynamical modeling and detailed analysis of heat capacity data. The thermal expansion measurements corroborate the previously suggested compensated AF ground state and the spin-flip metamagnetic transition at high fields and low temperatures. The thermal expansion and magnetostriction data manifest that the dominant contribution to the temperature and field evolution of measured change in speed of sound is coming from the change of the elastic moduli itself and the change of sample dimensions. The observed elastic properties regarding both --- topology of the magnetic phase diagram and temperature-field dependences of the observables are very well captured by the thermodynamical model based on the exchange-striction coupling in the antiferromagnetic Ising model.

Both the determined static and dynamic properties match the well-established magnetic phase diagram~\cite{Val, Val2} with a notable exception --- similarly to the transport properties~\cite{Val2} the anomalies in the temperature and field dependences of elastic moduli show a non-monotonous behavior along the second-order phase transition line. To understand this behavior, a detailed measurement of the heat capacity in the vicinity of the ordering temperatures under an applied magnetic field has been done and evaluated by means of critical behavior. The outcome confirms the 3D Ising model, at zero and low magnetic fields, with a crossover to the mean-field tricritical behavior at fields close to the tricritical point, where tricritical fluctuations dominate the temperature evolution of given property.

\section*{Acknowledgement}

Experiments were performed in MGML (mgml.eu), which is supported within the program of Czech Research Infrastructures (project no. LM2018096). We acknowledge the support by the Operational Program Research, Development and Education financed by European Structural and Investment Funds and the Czech Ministry of Education, Youth and Sports (Project MATFUN – CZ.02.1.01/0.0/0.0/15 003/0000487). The authors are indebted to Ross H. Colman for critical reading the manuscript and making language corrections.

\beginsupplement

\section*{\add{Supplemental Material}}

{\large{\center{Tricritical fluctuations and elastic properties of the Ising antiferromagnet UIrSi$_3$}}}
\center{T.N.~Haidamak, J.~Valenta, J.~Prchal, M.~Vališka, J.~Pospíšil, V.~Sechovský, J.~Prokleška}\\
{Charles University, Faculty of Mathematics and Physicss, Department of Condensed Matter Physics, Ke~Karlovu~5, Prague, Czech Republic}
\\
\center{A.A.~Zvyagin}\\
{B.~Verkin Institute for Low Temperature Physics and Engineering of the National Academy of Sciences of Ukraine,  47, Nauky ave., Kharkiv, 61103, Ukraine}\\
{V.N.Karazin Kharkiv National University, 4, Svobody sq., Kharkiv, 61022, Ukraine}
\\
\center{F.~Honda}\\
{Central Institute of Radioisotope Science and Safety Management Kyushu University Motooka 744, Fukuoka-Nishi, Fukuoka 819-0395, Japan}

\begin{figure}[h]
\begin{center}
\includegraphics[width=0.75\textwidth]{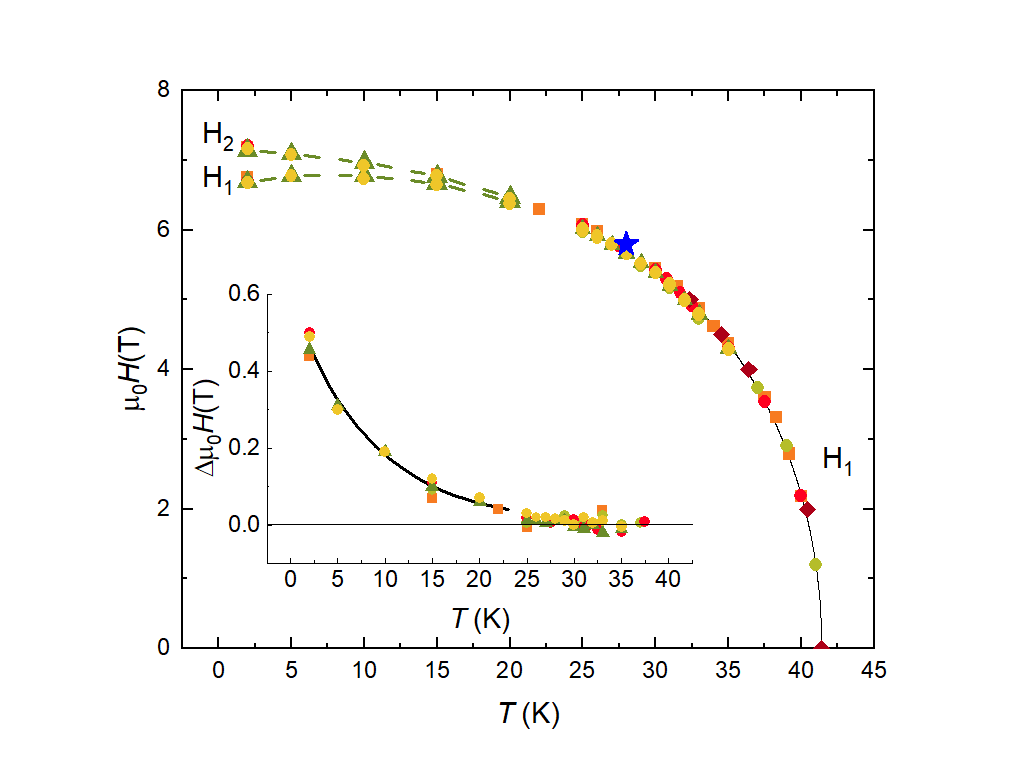}

\end{center}
\caption{(main plot) $B$--$T$ magnetic phase diagram based on the data in the main text (cf. earlier studies on the same compound --- Fig.~10 in~\cite{Val} and Fig.~10 in~\cite{Val2}). The solid line on the second order branch of $H_1$  is the result of the fit to eq.~\ref{para}, dashed lines (first order branch of $H_1$ and $H_2$) are guides to the eye, notations ($H_1,H_2$) reference to the discussion in the main text. Blue star indicate the position of the tricritical point. The inset shows the temperature dependence of the hysteresis, line is the result of the fit to the empirical exponential behaviour. \\
Data legend: 
$C_{33}$ red circle, $C_{44}$ orange square, $C_{11}-C_{12}$ green circle, magnetostriction $a$-axis green triangle, magnetostriction $c$-axis yellow circle, heat capacity dark red diamonds.
}
\label{MPD}
\end{figure}

\begin{figure}[h]
\begin{center}
\includegraphics[width=0.485\textwidth]{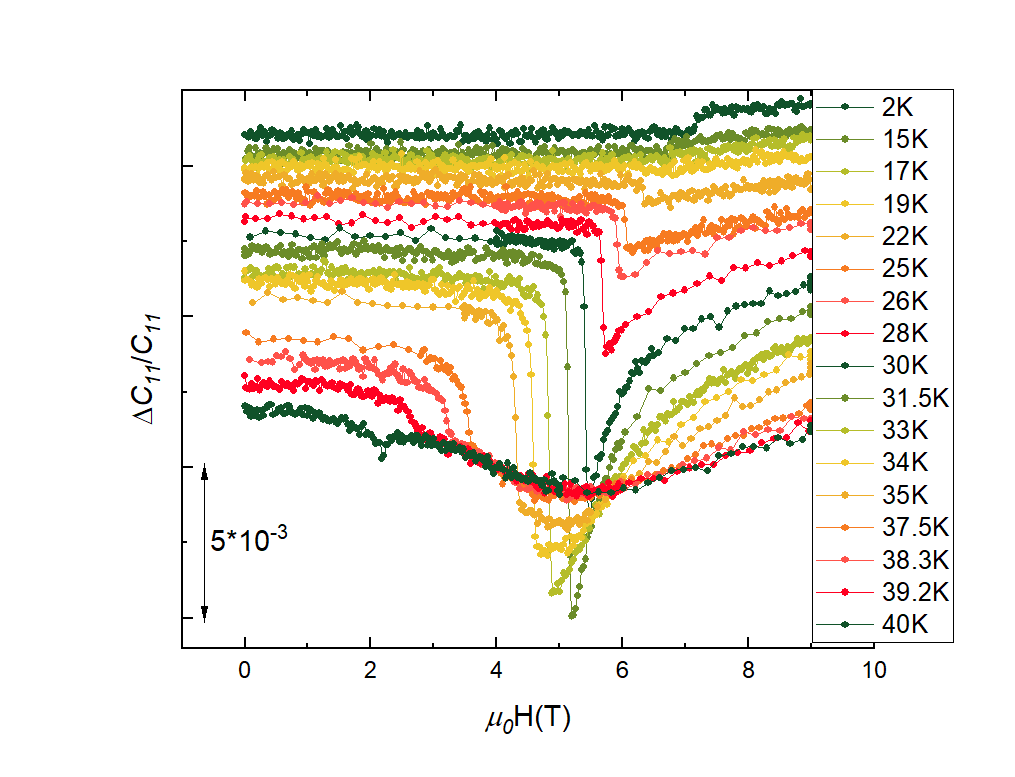}
\includegraphics[width=0.49\textwidth]{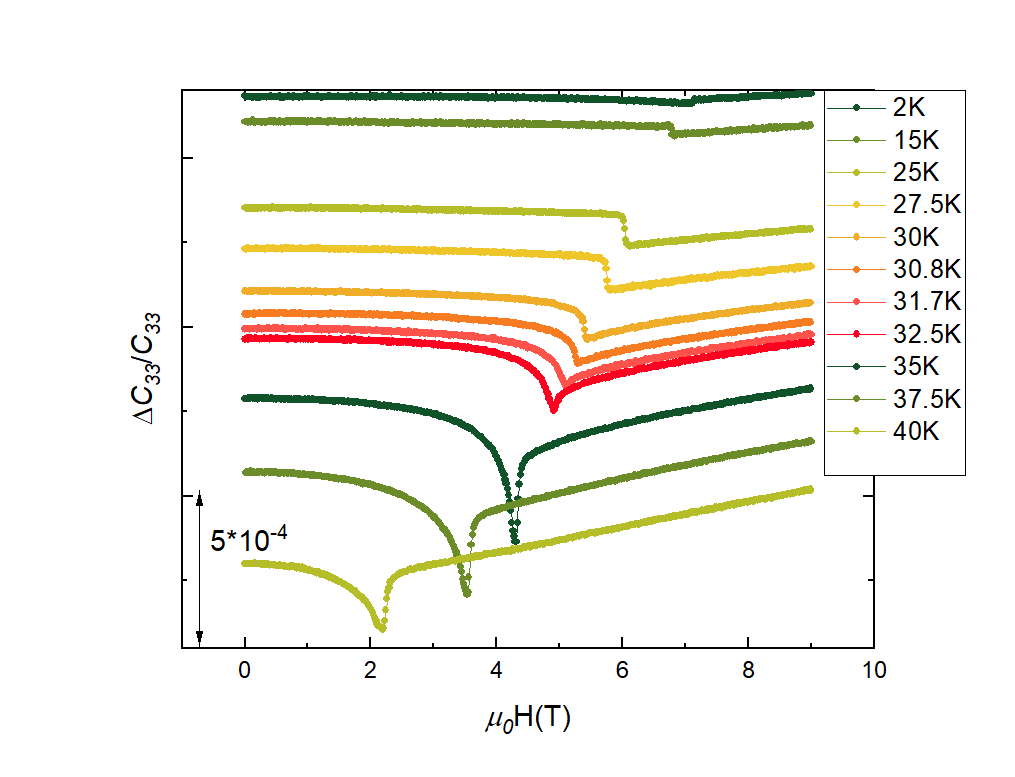}

\end{center}
\caption{ Isothermal field dependencies of $C_{11}$ and $C_{33}$ moduli for field applied along the $c$-axis, the curves are vertically shifted for better readability.
}
\label{figUZ_B_all}
\end{figure}

\begin{figure}[h]
\begin{center}
\includegraphics[width=0.485\textwidth]{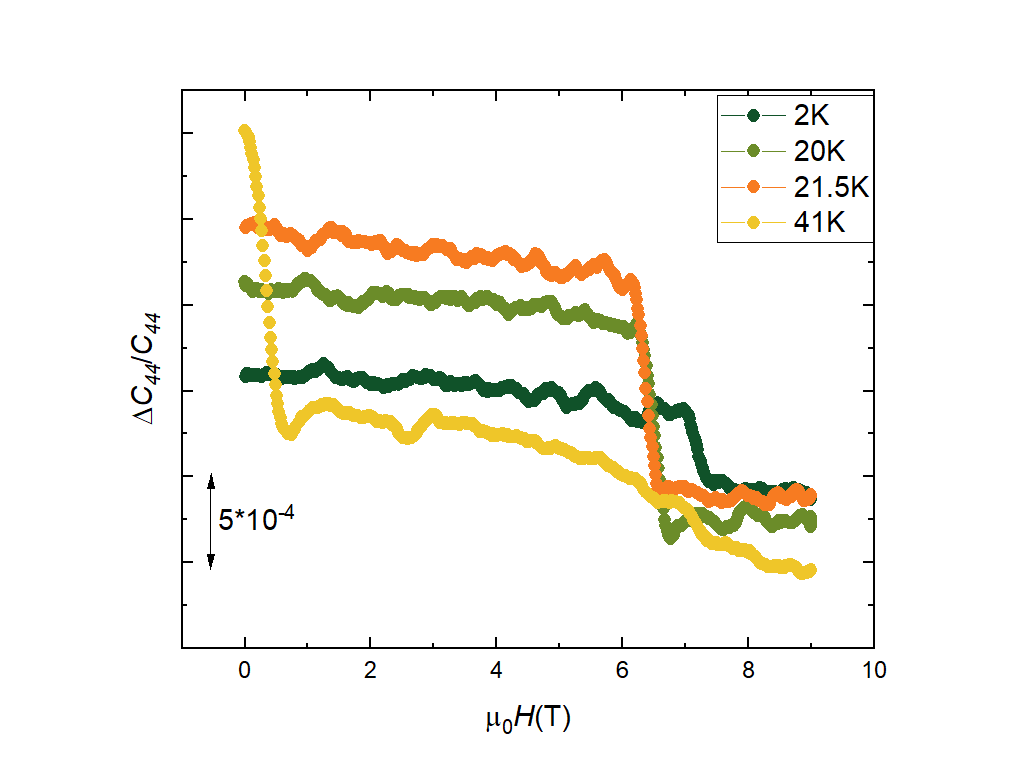}
\includegraphics[width=0.49\textwidth]{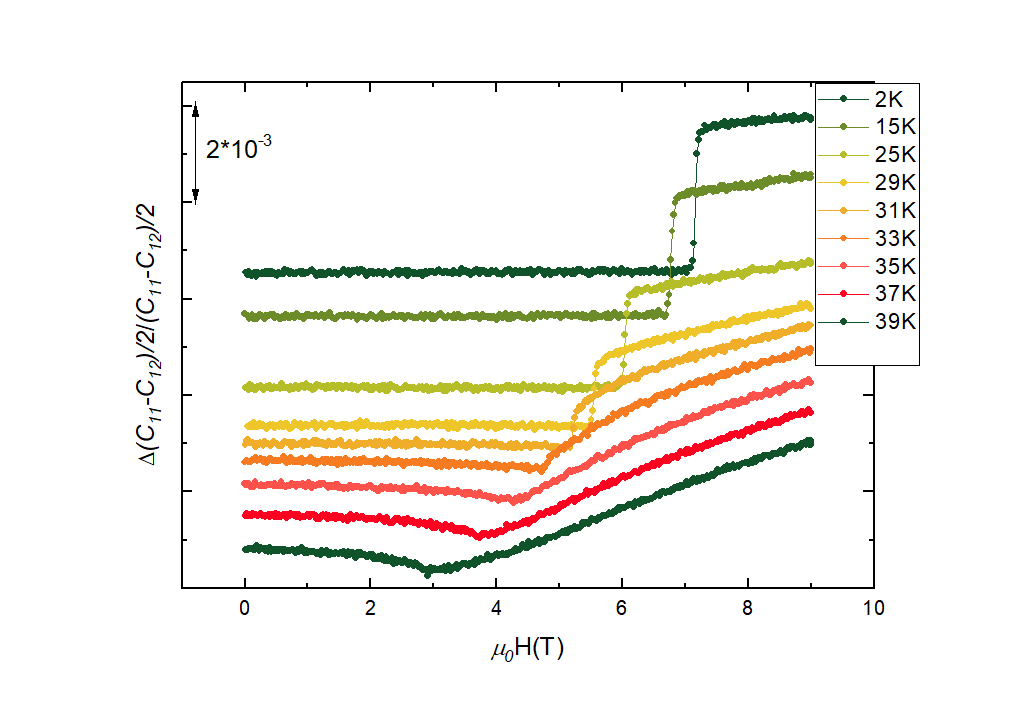}

\end{center}
\caption{Isothermal field dependencies of transverse modes $C_{44}$ (left) and $(C_{11}-C_{12})/2$ (right) for field applied along the $c$-axis, the curves are vertically shifted for better readability.
}
\label{figUZ_TM}
\end{figure}

\begin{figure}[h]
\begin{center}
\includegraphics[width=0.475\textwidth]{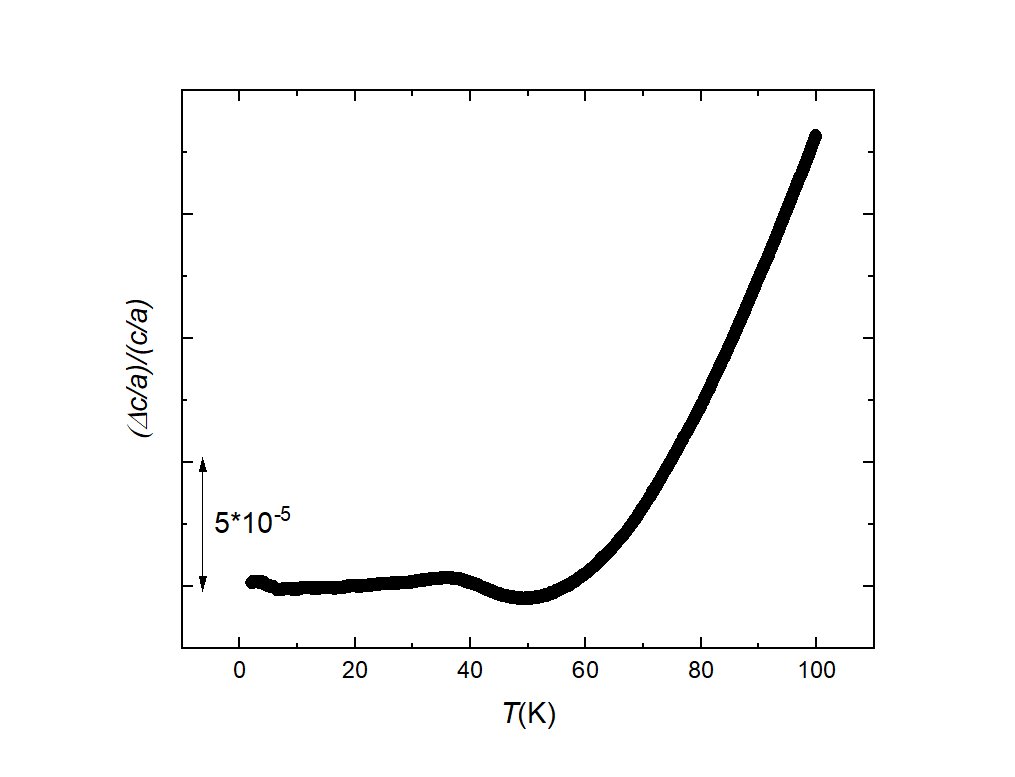}
\includegraphics[width=0.475\textwidth]{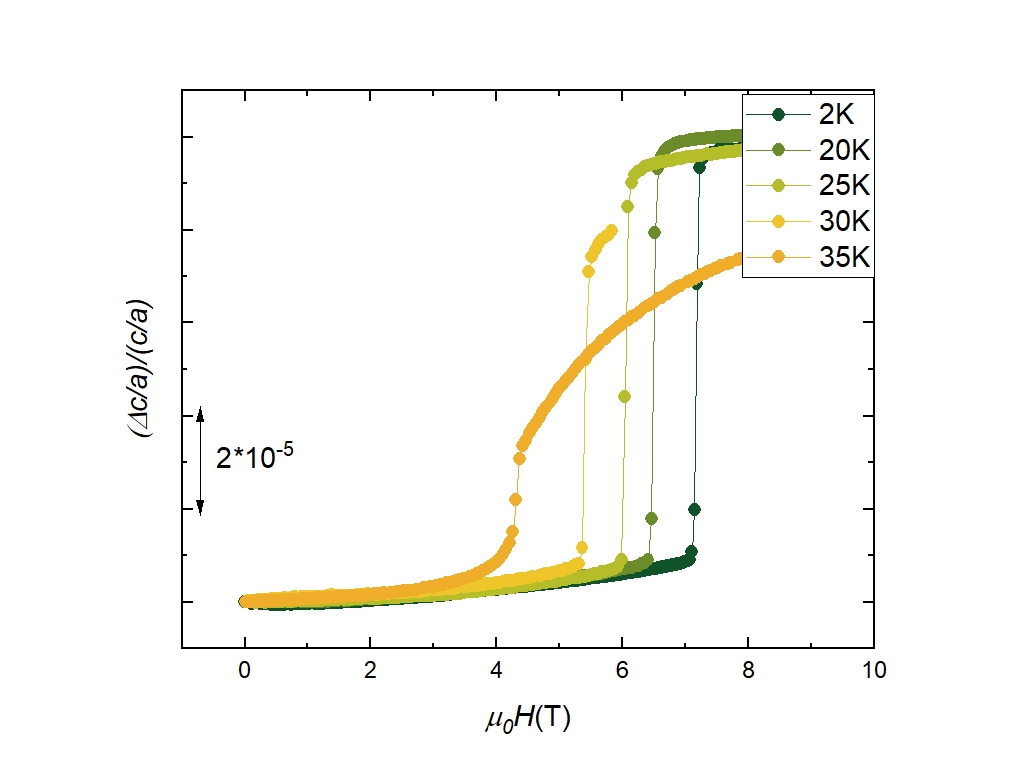}
\end{center}
\caption{The change of the tetragonality ($\frac{c}{a}$) as a function of temperature at zero magnetic field (left) and as a function of magnetic field at low temperatures (right).}
\label{tetragonality}
\end{figure}

\begin{figure}[h]
\begin{center}
\includegraphics[width=0.55\textwidth]{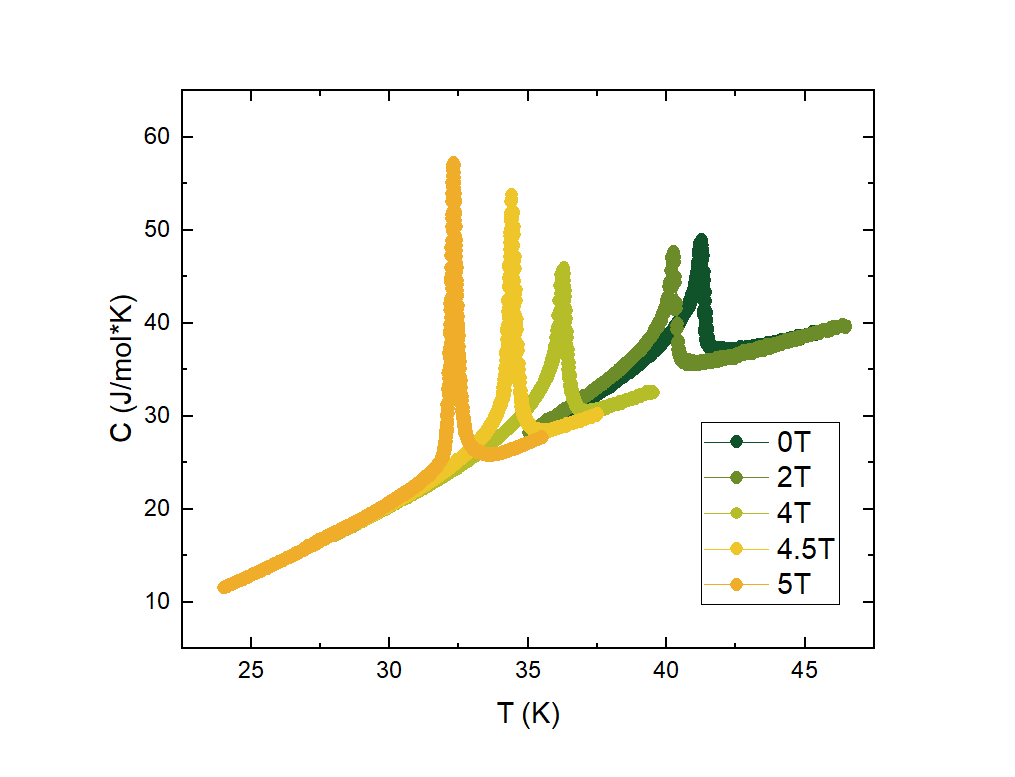}
\end{center}
\caption{ The temperature dependence of heat capacity measured for selected fields applied along the c-axis used for the evaluation of the  critical behaviour (cf.~Fig.~\ref{trivscri})}
\label{HC}
\end{figure}

\begin{figure}[h]
\begin{center}
\includegraphics[width=0.55\textwidth]{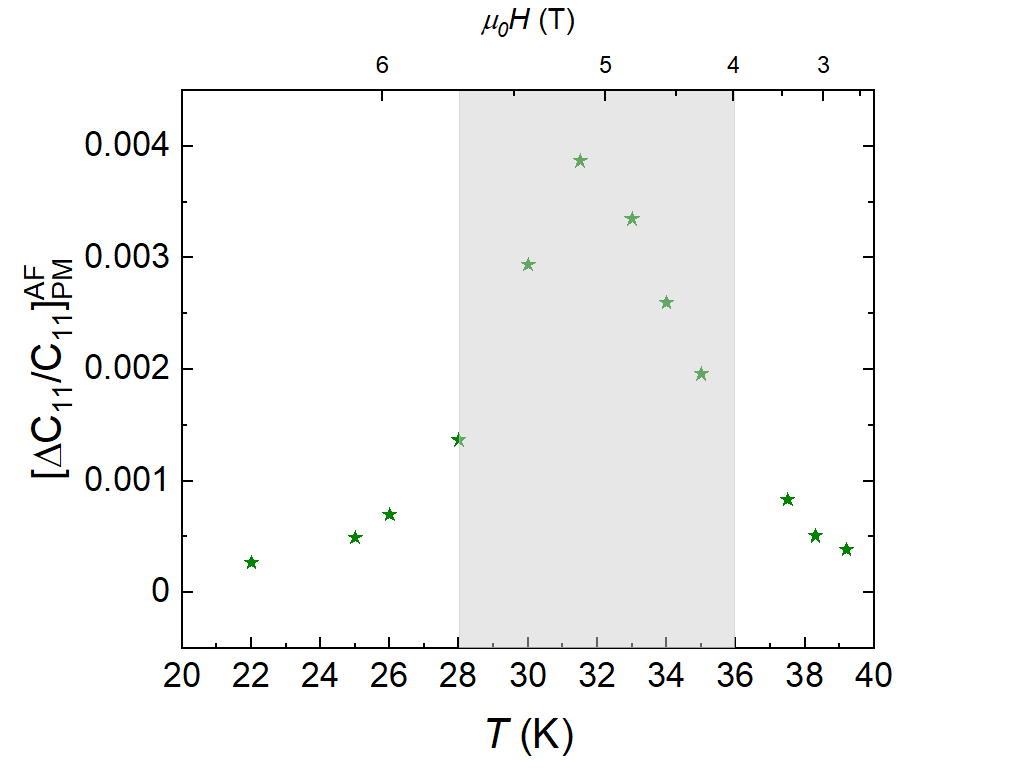}
\end{center}
\caption{The change in the relative elastic constant $\frac{\Delta C_{11}}{C_{11}}$ across the phase boundary, as evaluated from the~Figs.~\ref{figUZ_B} and~\ref{figUZ_B_all}. Shaded area indicates the prevailing tricritical character, as determined from critical behaviour in heat capacity (cf.~Fig.~\ref{trivscri})}
\label{transline}
\end{figure}

\end{document}